\documentclass{PoS}
\usepackage{epsfig}

\newcommand{\be}{\begin{equation}}
\newcommand{\ee}{\end{equation}}
\newcommand{\bea}{\begin{eqnarray}}
\newcommand{\eea}{\end{eqnarray}}
\newcommand{\ba}{\begin{array}}
\newcommand{\ea}{\end{array}}
\newcommand{\Tr}{\mathrm{Tr}}
\newcommand{\calR}{{\cal R}}
\newcommand{\bz}{{\bf z}}
\newcommand{\bp}{{\bf p}}
\newcommand{\bx}{{\bf x}}
\newcommand{\Fgen}{K}

\title{Thermal momentum distribution from shifted boundary conditions}

\ShortTitle{Thermal momentum distribution}

\author{\speaker{Leonardo Giusti}\\
        Dipartimento di Fisica, Universit\'a di Milano-Bicocca,\\
        Piazza della Scienza 3, I-20126 Milano, Italy\\
        E-mail: \email{Leonardo.Giusti@cern.ch}}

\abstract{At finite temperature the distribution of the total momentum is an 
observable characterizing the thermal state of a field theory, 
and its cumulants are related to thermodynamic potentials. 
In a relativistic system at zero chemical potential, for instance, 
the thermal variance of the total momentum is a direct measure of 
the entropy. We relate the generating function of the cumulants to 
the ratio of a path integral with properly shifted boundary 
conditions in the compact direction over the ordinary partition function. 
In this form it is well suited for Monte-Carlo 
evaluation, and the cumulants can be extracted straightforwardly. 
We test the method in the SU(3) Yang--Mills theory, and obtain the entropy 
density at three different temperatures.}

\FullConference{The XXIX International Symposium on Lattice Field Theory - Lattice 2011\\
July 10-16, 2011\\
Squaw Valley, Lake Tahoe, California}

\begin{document}

\section{Introduction}
\vspace{-0.375cm}

Thermal field theory is the theoretical tool for computing properties of 
matter at high temperatures and densities from first principles. It allows one, for instance, 
to determine the equation of state of Quantum Chromodynamics, which in turn is an essential 
ingredient to understand the properties of matter created in heavy ion collisions, and to 
model the behavior of hot matter in the early universe (for a review at this 
conference see Ref.~\cite{Levkova:2010}). Obtaining first-principles predictions from a 
thermal field theory is often challenging since it describes an infinite number of degrees 
of freedom subject to both quantum and thermal fluctuations.  Even though there are several 
established methods to compute the thermal properties of field 
theories~\cite{Engels:1981qx,Engels:1990vr,Endrodi:2007tq,Meyer:2009tq}, new theoretical 
concepts and more efficient computational techniques are still needed in many contexts
particularly when weak-coupling methods are inapplicable.

Recently Meyer and I proposed a new way to determine the equation of state of a thermal 
field theory, and it is the aim of this talk to discuss the results obtained in 
these papers~\cite{Giusti:2010bb,Giusti:2011kt}. We related the generating function 
of the cumulants of the total momentum distribution to a path integral with 
properly chosen shifted boundary conditions in the compact direction normalized to the 
ordinary one. By exploiting the Ward Identities (WIs) associated to the space-time 
invariances of the continuum theory, the cumulants can be related in a simple manner 
to thermodynamic potentials. In a relativistic theory at zero chemical potential, 
for instance, the variance of the momentum probability distribution measures the entropy 
of the system. 

Crucially the argument holds, up to harmless finite-size and discretization effects, 
also in a lattice box. The formulas are thus applicable at finite lattice spacing and volumes, 
where ratios of path integrals can be determined by \emph{ab initio} Monte Carlo computations. 
As a result the entropy density, the pressure and the specific heat can be obtained by studying 
the response of the system to the shift, and no additive or multiplicative 
ultraviolet-divergent renormalizations are needed in taking the continuum limit. We have numerically 
tested our proposal in the SU(3) Yang--Mills theory, for which the entropy density 
has been determined at three different temperatures.
\vspace{-0.25cm}

\section{Momentum distribution from shifted boundaries\label{sec:SyCPI}}
\vspace{-0.375cm}

For an Euclidean field theory at a temperature $T=1/L_0$, where $L_0$ is the 
length of the ``time'' direction, the relative contribution to the partition function of 
states with total momentum $\bp$ is given by 
\be\label{eq:SyCPI}
\frac{R({\bp})}{V} = 
\frac{{\rm Tr}\{e^{-L_0\, \hat {\rm H}}\, \hat{\rm P}^{({\bp})}\}}{{\rm Tr}\{e^{-L_0\, \hat {\rm H}}\}}
\ee
where $V=L^3$ is the spatial volume, and $L$ is the linear dimension in the three 
spatial directions. The trace 
in Eq.~(\ref{eq:SyCPI}) is over all the states of the Hilbert space, $\hat{\rm
  P}^{({\bp})}$ is the projector onto those states with total momentum
$\bp$, and $\hat {\rm H}$ is the Hamiltonian of the theory. If we 
introduce the partition function
\be
Z(\bz)=\Tr\{e^{-L_0\, \hat {\rm H}}\, e^{i\hat\bp\bz}\}
\ee
in which states of momentum $\bp$ are weighted by a phase $e^{i\bp\cdot\bz}$, 
and we use the standard group theory machinery 
(see Refs.~\cite{DellaMorte:2008jd,DellaMorte:2010yp} for a detailed discussion 
of this point) it easy to show that 
\be\label{eq:bella2}
R({\bp}) = \frac{1}{Z}\, \int d^3{\bz}\, e^{-i{\bp}\cdot{\bz}}\, Z(\bz)\; .
\ee
The generating function $\Fgen({\bz})$ of the cumulants is defined as  
\be\label{eq:freeE}
e^{-\Fgen({\bz})} = 
\frac{1}{V} \sum_{\bp} e^{i{\bp}\cdot{\bz}} \, R({\bp})\;,
\ee
and the cumulants are given by 
\be\label{eq:cum1}
k^{_V}_{\{2 n_1, 2 n_2, 2 n_3\}} = (-1)^{n_1+n_2+n_3+1}\, 
\frac{\partial^{2n_1}}
{\partial z_1^{2n_1}} \frac{\partial^{2n_2}}{\partial z_2^{2n_2}}
\frac{\partial^{2n_3}}{\partial z_3^{2n_3}}
\frac{\Fgen({\bz})}{V}\Big|_{\bz=0}\; ,
\ee
where they have been normalized so as to have a finite limit when $V\rightarrow\infty$. 
The shifted partition function $Z(\bz)$ can be expressed as an Euclidean path 
integral with the field satisfying the boundary conditions
\be
\phi(L_0,{\bx})=\pm\phi(0,{\bx}+{\bz})
\ee
with the $+$ ($-$) sign for bosonic (fermionic) fields respectively. 
From Eqs.~(\ref{eq:bella2}) and  (\ref{eq:freeE}), the generating function can 
thus be written as the ratio of partition functions
\be\label{eq:exp-F}
e^{-\Fgen({\bz})} = \frac{Z(\bz)}{Z}\;,
\ee 
i.e. two path integrals with the same action but different boundary 
conditions, and the cumulants can be obtained by 
deriving it with respect to the shift parameter $z$ an appropriate 
number of times. Being the cumulants the connected correlation functions
of the charges associated to the translational invariance of the theory, 
the functional $\Fgen({\bz})$ and the momentum distribution 
$R({\bp})$ are ultraviolet finite, see Refs.~\cite{Giusti:2010bb,Giusti:2011kt}
for a comprehensive discussion of this point.
\vspace{-0.125cm}

\subsection{Extension to the lattice}
\vspace{-0.125cm}

When defined on a lattice, the theory is invariant under a discrete
subgroup of translations and rotations only, the momenta are quantized, 
and the continuum WIs are broken by discretization effects. Generic lattice 
definitions of the energy-momentum tensor, as well as the corresponding
charges, require ultraviolet renormalization. It is still possible, however, 
to factorize the Hilbert space of the lattice theory in sectors with definite 
conserved total momentum. The formula for the lattice projector 
is given by
\be
\hat{\rm P}^{({\bp})} = \frac{a^3}{V} \sum_{\bz}\, 
e^{-i\, {\bf p}\cdot {\bf z}}\, e^{i\,\hat\bp\bz}
\ee
where the sum is over all the lattice points. Since only physical states 
contribute to the 
symmetry constrained path integrals in Eq.~(\ref{eq:SyCPI}) when $L_0\neq 0$,
the lattice momentum distribution $R({\bp})$ is expected to converge to the 
continuum universal one without the need for any ultra-violet renormalization.   
The definition of the cumulants in Eq.~(\ref{eq:cum1}) is thus applicable at 
finite lattice spacing, provided the derivatives are replaced with their discrete 
counterpart, and no additive or multiplicative ultraviolet-divergent 
renormalization is needed for taking the continuum limit.

\subsection{Extension to other symmetries}
\vspace{-0.125cm}

The factorization of the Hilbert space into sectors whose states 
have definite transformation properties under a symmetry of 
the lattice theory is more generally applicable than just for 
translations. For a generic discrete group, the cubic rotations 
for instance, the projector onto the states  which transform 
as an irreducible representation $\mu$ is given by
\[
\hat{\rm P}^{(\mu)} = \frac{n_\mu}{g} \sum_{i=1}^{g}\, 
\chi^{(\mu)*}(\calR_i)\,\hat \Gamma(\calR_i)\; ,\qquad
\hat \Gamma(\calR_i)\, |\phi\rangle =|\phi^{\calR_i}\rangle \; ,
\]
where $n_\mu$ is the dimension of the representation,
$g$ the order of the group, $\chi^{(\mu)}(\calR_i)$ the character 
of the irreducible representation for the group element $\calR_i$, and 
$\hat \Gamma(\calR_i)$ is the representation of the group element
onto the Hilbert space. The generalization to a continuum symmetry, 
such as for instance the baryonic number, is straightforward.
The connected correlation functions of the corresponding charges 
can be extracted from the  
symmetry constrained path integrals, defined analogously to 
Eq.~(\ref{eq:SyCPI}), by studying the response of the system
to the properly chosen twist in the boundary 
conditions \cite{DellaMorte:2010yp}.
\vspace{-0.375cm}

\section{Continuum Ward identities and connection to thermodynamics}
\vspace{-0.375cm}

In the continuum theory the invariance under space-time translations implies the WIs
\be\label{eq:WIstd1}
\epsilon_\nu\, \langle \partial_\mu T_{\mu\nu}(x)\, O_1 \dots O_n \rangle = -  
\sum_{i=1}^{n}\, \left\langle O_1 \dots \delta^x_\epsilon O_i \dots O_n \right\rangle\; , 
\ee
where $O_i$ is a generic local field, $\delta^x_\epsilon O_i$ is its variation 
under the local transformation parameterized by $\varepsilon_\nu(z) = \epsilon_\nu \delta^{(4)}(z-x)$, 
and $T_{\mu\nu}$ is the energy-momentum tensor (see Ref.~\cite{Giusti:2011kt} for 
unexplained notation). By choosing $\epsilon_\nu = \delta_{\nu k} \epsilon_k$ (no summation 
over $k$), $\widetilde{\overline{T}}_{00}(x^1)$ as interpolating operator, 
the WI in Eq.~(\ref{eq:WIstd1}) and translational invariance lead to   
\be\label{eq:00kkJ12}
\partial^{x^1}_0 \langle \widetilde{\overline{T}}_{00}(x^1)\, 
\widetilde{\overline{T}}_{0k}(x^2)\rangle_{c} =
\partial^{x^2}_k \langle \widetilde{\overline{T}}_{kk}(x^2)\, 
\widetilde{\overline{T}}_{00}(x^1)\rangle_{c}\; , \qquad 
\widetilde{\overline{T}}_{\mu\nu}(x) = \int \Big[\prod_{\rho\neq 0,k} d x_\rho\Big]\, T_{\mu\nu}(x)\; .
\ee
By choosing $\epsilon_\nu = \delta_{\nu 0} \epsilon_0$,
$\widetilde{\overline{T}}_{0k}(x^2)$ as interpolating operator, and thanks to 
translational invariance and the symmetry of $T_{0k}$, 
the WI in Eq.~(\ref{eq:WIstd1}) gives 
\be\label{eq:00kkJ2}
\partial^{x^1}_0 \langle \widetilde{\overline{T}}_{00}(x^1)\, 
\widetilde{\overline{T}}_{0k}(x^2)\rangle_{c} =
\partial^{x^2}_k \langle \widetilde{\overline{T}}_{0k}(x^1)\, 
\widetilde{\overline{T}}_{0k}(x^2)\rangle_{c}\; .
\ee
By putting Eqs.~(\ref{eq:00kkJ12}) and (\ref{eq:00kkJ2}) together
we arrive to 
\be
\partial^{x^2}_k
\Big\{\langle \widetilde{\overline{T}}_{0k}(x^2)\, 
\widetilde{\overline{T}}_{0k}(x^1)\rangle_{c} -
\langle \widetilde{\overline{T}}_{kk}(x^2)\, 
\widetilde{\overline{T}}_{00}(x^1)\rangle_{c}\Big\} = 0\; . 
\ee
By integrating in $x^2_k$, while  keeping all insertions at a physical 
distance ($x_0^i$ all different), we obtain
\be\label{eq:final0}
\hspace{-0.175cm}\Big\{ 
\langle {\overline T}_{0k}(x_0^1)\, {\overline T}_{0k}(x_0^2) \rangle_{c} - 
\langle {\overline T}_{00}(x_0^1)\,  {\overline T}_{kk}(x_0^2)  \rangle_{c} \Big\} 
=
L_k^2\; 
\Big\{\langle \widetilde{\overline{T}}_{0k}(x^1)\, 
\widetilde{\overline{T}}_{0k}(x^2) \rangle_{c} - 
\langle \widetilde{\overline{T}}_{00}(x^1)   \widetilde{\overline{T}}_{kk}(x^2)\, 
\rangle_{c} \Big\} \; , 
\ee
where ${\overline T}_{0k}(x_0)= \int d^3 x\, T_{0k}(x)$. If we remember that 
in the Euclidean the momentum operator maps to  
$\hat p_k \rightarrow -i {\overline T}_{0k}$, the pressure maps to 
$ p = \langle T_{kk} \rangle$, and if we note that the r.h.s. of 
Eq.~(\ref{eq:final0}) vanishes in the thermodynamic limit, we arrive to
infinite-volume relation
\be
k_{\{0,0,2\}} = T^2 \frac{\partial}{\partial T} p\; . 
\ee
Combined with the infinite-volume WI $s=\frac{\partial}{\partial T} p$, where 
$s$ is the entropy density, it leads to \cite{Giusti:2010bb,Giusti:2011kt} 
\be\label{eq:final1}
s = - \frac{1}{T^2}\,\lim_{V\rightarrow \infty} \frac{1}{V} \frac{d^2}{d z^2}
\ln{Z(\{0,0,z\})}\Big|_{z=0}\; .
\ee
By following an analogous derivation, it is possible to show that
the specific heat is given by~\cite{Giusti:2011kt}
\be
c_v =  \lim_{V\rightarrow \infty} \frac{1}{V}\left[
\frac{1}{3 T^4}\frac{d^4}{d z^4} + \frac{3}{T^2} \frac{d^2}{d z^2}\right]
 \ln{Z(\{0,0,z\})}\Big|_{z=0}\; . 
\ee  
The last two equations make clear that the response of the partition function 
to the shift $z$ is governed by basic thermodynamic properties of the system, 
and that the potentials entering the equation of state of the thermal theory 
can be extracted by rather simple formulas. Crucially the convergence to the 
thermodynamic limit is exponential in $ML$, where $M$ is the lightest screening 
mass of the theory~\cite{Giusti:2011kt}. 
\vspace{-0.375cm}

\section{Numerical computation}
\vspace{-0.375cm}

We have tested the numerical feasibility of the 
computational strategy presented above in the SU(3) 
Yang--Mills theory. The lattice theory is set up on a 
finite four-dimensional lattice with a spacing $a$ 
and periodic boundary conditions in the space directions. It is discretized 
by the standard plaquette Wilson action
\be
S[U] = \frac{6}{g_0^2}\, \sum_{x} \sum_{\mu<\nu} 
\left[1 - \frac{1}{3}{\rm Re}\Tr\Big\{U_{\mu\nu}(x)\Big\}\right]\; ,
\ee
where the trace is over the color index, and $g_0$ is the bare coupling constant. 
The plaquette $U_{\mu\nu}(x)$ and the path integral $Z$ are defined as usual.
The most straightforward way for computing the cumulant generator is to rewrite it as 
\be
\frac{Z({\bf z})}{Z} = \prod_{i=0}^{n-1}
\frac{{\cal Z}({\bf z},r_i)}{{\cal Z}({\bf z},r_{i+1})}\; , 
\ee
where a set of $(n+1)$ systems is designed so that the relevant phase spaces of successive path 
integrals overlap, and that ${\cal Z}({\bf z},r_0)=Z({\bf z})$ and 
${\cal Z}({\bf z},r_n)=Z$.  The path integrals of the interpolating 
systems that we have implemented are defined as  
\be
Z({\bf z},r) = \int D U\, D U_{4,(L_0/a)-1}\,                    
\; e^{-\overline S[U,U_4,r]}\; , 
\ee
where $U_{4,(L_0/a)-1}$ is an extra ($5^{th}$) temporal link assigned to each 
point of the $((L_0/a)-1)$ time-slice. The action of each system is  
\be
\overline S[U,U_4,r] = S[U] + \frac{\beta}{3}(1-r)
\sum_{{\bf x}, k} 
{\rm Re}\Tr\Big\{U_{0 k}((L_0/a)-1,{\bf x}) - U_{4 k}((L_0/a)-1,{\bf x}) \Big\}\; , 
\ee
with the extra space-time plaquette given by
\be
U_{4 k}((L_0/a)-1,{\bf x}) = U_4((L_0/a)-1,{\bf x})\, 
U_k(0,{\bf x}+{\bf z})\, 
U^\dagger_4((L_0/a)-1,{\bf x}+\hat k)\,
             U^\dagger_k((L_0/a)-1,{\bf x})\; . 
\ee
If the "reweighting'' observable 
\be
O[U,r_{i+1}] = 
e^{{\overline S}[U,U_4,r_{i+1}]-{\overline S}[U,U_4,r_i]}
\ee
is defined, then 
\be
\frac{{\cal Z}({\bf z},r_i)}{{\cal Z}({\bf z},r_{i+1})} = 
\left\langle\, O[U,r_{i+1}]\, \right\rangle_{r_{i+1}} \; , 
\ee
and the entropy can be computed as
\be\label{eq:Ssn} 
s = -\frac{2}{{\bf z}^2  T^2 L^3} \sum_{i=0}^{n-1} 
\ln{\left[\frac{{\cal Z}({\bf z},r_{i})}{{\cal Z}({\bf z},r_{i+1})}\right]}\; ,
\ee
with $\bz=(0,0,n_z a)$ and with the integer $n_z$ being kept fixed when $a\rightarrow 0$. 
\begin{table}[t!]
\begin{center}
\begin{tabular}{llcccccc}
\hline\\[-0.325cm]
Lat        &$6/g^2_0$&$L_0/a$&$L/a$& $\frac{1}{n}(\frac{L}{a})^3$&$r_0/a$&$K({\bz})$ & 
$\frac{2 \Fgen(\bz)}{{|{\bz}|}^2 T^5 L^3}$\\ 
\hline
${\rm A}_1$   &$5.9$  & $4$&$12$& 2 & $4.48(5)$&$17.20(11)$&$5.10(3)$ \\
${\rm A}_{1a}$&$5.9$  & $4$&$16$& 2 & $4.48(5)$&$40.71(15)$&$5.089(19)$ \\
${\rm A}_2$   &$6.024$& $5$&$16$& 2 & $5.58(6)$&$13.05(10)$&$4.98(4)$ \\
${\rm A}_3$   &$6.137$& $6$&$18$& 3 & $6.69(7)$&$7.32(8)$  &$4.88(6)$ \\
${\rm A}_4$   &$6.337$& $8$&$24$& 4 & $8.96(9)$&$4.32(16)$ &$5.12(19)$\\
${\rm A}_5$   &$6.507$&$10$&$30$& 5 & $11.29(11)$&$2.62(17)$ &$4.9(3)$\\
\hline
${\rm B}_1$   &$6.572$& $4$&$12$& 2 & $12.28(12)$&$22.22(11)$&$6.58(3)$ \\
${\rm B}_{1a}$&$6.572$& $4$&$16$& 2 & $12.28(12)$&$53.47(16)$&$6.684(20)$ \\
${\rm B}_2$   &$6.747$& $5$&$16$& 2 & $15.34(15)$&$17.11(15)$&$6.53(6)$ \\
${\rm B}_3$   &$6.883$& $6$&$18$& 3 & $18.14(18)$&$9.61(9)$  &$6.40(6) $\\
${\rm B}_4$   &$7.135$& $8$&$24$& 4 & $24.5(3)$&$5.42(17)$ &$6.42(20)$\\
${\rm B}_5$   &$7.325$&$10$&$30$& 5 & $30.7(4)$&$3.32(18)$ &$6.1(3)$\\
\hline
${\rm C}_1$   &$7.234$& $4$&$16$& 4 & $27.6(3)$&$57.44(25)$&$7.18(3)$ \\
${\rm C}_2$   &$7.426$& $5$&$20$& 5 & $34.5(4)$&$36.5(4)$&$7.13(8)$ \\
${\rm C}_3$   &$7.584$& $6$&$24$& 4 & $41.4(5)$&$24.7(4)$&$6.94(12)$ \\
\hline
\end{tabular}
\caption{Lattice parameters and numerical results with $\bz=(0,0,2a)$.\label{tab:lattices}}
\end{center}
\end{table}
\vspace{-0.375cm}

\section{Results and conclusions}
\vspace{-0.375cm}

\begin{figure}[!t]
\begin{center}
\vspace{0.25cm}

\includegraphics[width=8.0cm]{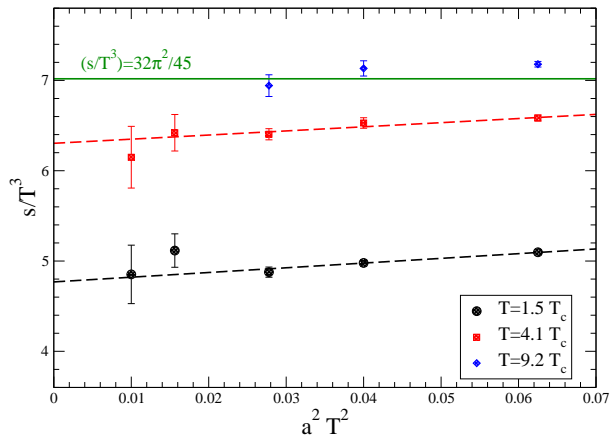}
\caption{Scaling behavior of $s/T^3$, 
see Eq.~(\protect\ref{eq:Ssn}). The Stefan-Boltzmann value 
is also displayed.\label{fig:conti}}
\end{center}
\end{figure}

Our goal is to show that the entropy density can be computed in the thermodynamic 
and continuum limit by using Eq.~(\ref{eq:Ssn}). To this aim we 
have calculated the entropy at three temperatures, 1.5, 4.1 and
9.2$T_c$, see Table \ref{tab:lattices} for the numerical results.
The update algorithm used is a standard combination of heatbath and
over-relaxation sweeps. The only changes over the standard algorithm 
reside (a) in the computation of
the extra ``staples'' that determine the contribution to the action 
$\overline S[U,U_4,r]$ of a given
link variable $U_\mu(x)$, and (b) in the more frequent updating of the
two time-slices on which the observable $O[U,r_{i+1}]$ has its support.
The bare coupling $g_0^2$ was tuned using the data
of Ref.~\cite{Necco:2001xg} in order to match lattices of different
$L_0/a$ to the same temperature.  Motivated by a study of the free
case, we chose $n_z=2$. A comparison of the values of $s$ for the 
lattices $A_1$ and $A_{1a}$, and ${\rm B}_1$ and ${\rm B}_{1a}$
indicates that finite-volume effects are very small or negligible
within our statistical errors for these lattices. The behavior of 
the entropy density 
as a function of the lattice spacing is displayed in 
Figure~\ref{fig:conti}. Cutoff effects are clearly quite mild. 
For illustration in the same plot we show a linear 
extrapolation in $a^2$ for the two lower temperatures where we have 
enough points. The corresponding intercepts are $4.77(8)$ and $6.30(9)$ at $1.5$ 
and 4.1$T_c$ respectively, where the errors are statistical only. 
These results prove the numerical feasibility of the strategy discussed 
in the previous sections for computing the entropy 
density of the SU(3) Yang--Mills theory. It is also worth noting that, 
even though the statistical errors for the most expensive runs at 
$L_0/a=8,10$ are still quite large for a solid continuum extrapolation, 
the continuum-limit results at $1.5$ and 4.1$T_c$ are compatible with 
the best published ones~\cite{Boyd:1996bx,Namekawa:2001ih}.\\[-0.375cm]

I thank Harvey B. Meyer for the intense and productive
collaboration over the last year which led to the results 
discussed here. Many thanks to the organizers of the conference 
for their work, and for giving us the possibility to discuss
physics in such a wonderful place.

\end{document}